# OPTICAL DEEP LEARNING NANO-PROFILOMETRY

Jinlong Zhu[1,†], Yanan Liu[2,†], Sanyogita Purandare[1], Jian-Ming Jin[2], Shiyuan Liu[3,*], and Lynford L. Goddard[1,*]

[1]Photonic Systems Laboratory, Department of ECE, UIUC, Urbana, IL, USA
[2]Center for Computational Electromagnetics, Department of ECE, UIUC, Urbana, IL, USA
[3]State Key Laboratory of Digital Manufacturing Equipment and Technology, HUST, Wuhan, China
†Equal contribution.
*Corresponding authors: shyliu@hust.edu.cn; lgoddard@illinois.edu.

## ABSTRACT

Determining the dimensions of nanostructures is critical to ensuring the maximum performance of many geometry-sensitive nanoscale functional devices. However, accurate metrology at the nanoscale is difficult using optics-based methods due to the diffraction limit. In this article, we propose an optical nano-profilometry framework with convolutional neural networks, which can retrieve deep sub-wavelength geometrical profiles of nanostructures from their optical images or scattering spectra. The generality, efficiency, and accuracy of the proposed framework are validated by performing two different measurements on three distinct nanostructures. We believe this work may catalyze more explorations of optics-based nano-metrology with deep learning.

## 1 Introduction

With the continuous development of innovative materials and advanced nanoscale fabrication techniques, a wide range of functional nanostructures can be fabricated today to address various scientific and engineering challenges across many fields, e.g., integrated photonic devices, metamaterials, and semiconductor transistors, to name a few. More specific examples include nanoscale scatterers for tuning a whispering gallery-mode micro-toroid cavity to operate at non-Hermitian spectral degeneracies [1], phase-gradient metasurfaces for controlling the propagation and coupling of waveguide modes [2], and the sub-10-nm fin-based field-effect transistor and gate all around devices used in the semiconductor industry [3]. However, as the aforementioned devices shrink in size and become more complicated in shape, dimensional metrology becomes increasingly important because the high performance of nanoscale devices is usually accompanied with an ultra-high sensitivity to their geometrical dimensions. For example, the gate-all-around transistors at the 7-nm node and beyond may not work properly when their critical dimensions deviate by 20% from the nominal design values. Hence, it is vital to accurately determine the geometrical dimensions of functional nanostructures for quality control purposes and to ensure their maximum performance.

Transmission electron microscopy (TEM) and scanning electron microscopy (SEM) enable direct imaging of entire nanostructures with single-digit nanometer resolutions due to the ultra-small de Broglie wavelength of electrons. However, they suffer from destructive sample preparation and low integratability [4, 5]. Scanning tunneling microscopy and atomic force microscopy could provide an even higher resolution than SEM, but are inherently inefficient due to the requirement of serial scanning [6, 7]. Super-resolved fluorescence microscopy offers significant advantages in particle position measurements by localizing the emitting fields [8, 9], but the artificially introduced fluorescence dyes are incompatible with solid nanostructures. Recent developments in near-field related super-resolution imaging techniques have shattered the diffraction limit by up to two orders of magnitude for certain objects [10, 11], but the high complexity and ultra-small field-of-view make them impractical for many applications. Far-field-based techniques such as optical profilometry and optical diffractive tomographic microscopy [12] are fast and can be easily integrated into a fabrication chain, but they have limited transverse resolutions and are usually used to measure height variations only. Optical scatterometry, which works by matching the experimental and simulated patterns using either a look-up-table or nonlinear regression [13, 14], enables a nanoscale measurement accuracy, but the intrinsic parameter coupling may hinder the accurate reconstruction of the critical dimensions in the devices, and yet there is no clue on the table to conquer this drawback [15–19]. Although there are efforts to improve optical scatterometry by introducing an artificial neural network [20] and a support vector machine [15], the limitation of dealing with low-dimensional datum in conventional machine learning makes it more suitable to the identification of nanoscale profile.

With the recent advances in computer vision and machine learning, convolutional neural networks (CNNs) have been shown to excel at many image processing tasks such as classification [21], recognition [22], and segmentation [23]. More recently, CNNs have been applied to parameter estimation from astrology images [24] and resolution enhancement and phase retrieval in bio-medical imaging [25–31]. In this study, we propose a CNN-assisted optical nano-profilometry framework that enables the non-destructive, non-contact, and accurate geometrical dimension reconstruction of nanostructures with deep subwavelength features. Different from existing deep learning based super-resolution schemes, we use CNNs to extract geometrical dimensions directly. We demonstrate the effectiveness and generality of the proposed framework with two drastically different optical modalities, i.e., optical bright-field microscopy (BFM) and ellipsometry; and three different measurands, i.e., a NIST RM 8820 [32] artifact, a nanoimprinted nanowire array, and a single dynamic random-access memory (DRAM) transistor at an advanced technology node. These types of nanostructures are used extensively in many fields including semiconductor industry, integrated photonics, metamaterials, one-dimensional photonic crystals, biosensors, and neuromorphic chips. The results show that the proposed framework can achieve dimension reconstruction of up to 1-nm scale accuracy with the field of view at 10,000 $\mu m^2$ scale. Further, the reconstruction accuracy of individual dimensions can be controlled via tuning the weights in the loss function of the CNN, offering a degree-of-freedom in investigating the most critical set of parameters that govern the device performance. Such a feature in the method overcomes the issue of intrinsic parameter coupling from deterministic reconstruction in conventional pattern-matching-based nanometrology techniques, where the critical set of parameters may not be accurately reconstructed because of their coupling with influential but less significant parameters in the inverse reconstruction process [33]. Although BFM and ellipsometry are our focus in this paper, we expect the proposed framework to be capable of working with other modalities, e.g., dark-field [34], optical coherence tomography [35, 36], diffractive tomography microscopy [37], and diffractive phase microscopy [38–40], for nanoscale metrology, provided that the corresponding imaging process can be numerically modeled for training data generation.

## 2 Methods

### 2.1 The proposed framework

The proposed CNN metrology framework is shown in Fig. 1a, which consists of four steps: training data generation, CNN training, optical measurement, and dimension reconstruction. The first step is training data generation through simulation. The simulation method has to match the measurement system and be able to model the physics of the optical modality. In our examples, we use Fourier optics-based simulation [41] for the BFM, and rigorous coupled wave analysis (RCWA) [42] for the ellipsometry. For data generation, we randomly sample the geometric space of the measurand and use the appropriate simulation method to generate the image (for BFM) or the wavelength-resolved or angle-resolved spectrum (for ellipsometry) for each data sample. The generated optical data are divided into the training set for optimizing the CNN parameters, the validation set for model selection, and the test set for model evaluation. The CNN models the inverse process of the physics-based simulation, where the input is the image or spectrum and the output is the reconstructed dimensions. After the training is completed, the CNN model can be directly used with measurement data, which are either raw intensity images or scattering spectra, obtained by a bright-field microscope or a Mueller matrix ellipsometer (see more details in Sec. 2C), respectively.

### 2.2 CNN model training

The CNN model can be represented mathematically as a general composite function given by

$$\mathbf{y} = F(\Theta, \mathbf{x}) = f_1\left(\Theta_1, f_2\left(\Theta_2, f_3\left(...f_n\left(\Theta_n, \mathbf{x}\right)...\right)\right)\right) \tag{1}$$

where $f_1, f_2, ..., f_n$ are the layers of the network and can be operations such as convolution, pooling, batch normalization, and dropout. $\Theta$ denotes the network parameters to be optimized. Our models follow the VGGNet [21] architecture, with successive modules of convolution–batch-normalization–pooling–dropout layers followed by fully-connected layers at the output. An illustration is shown Fig. 1c whereas the details of the models can be found in Table S1 in Supplement 1. The objective of training is given as

$$\min_{\Theta} \frac{1}{|N|} \sum_{n=1}^{N} \sum_{i=1}^{M} \omega_{yi} \times \left(\hat{y}_i^{(n)} - y_i^{(n)}\right). \tag{2}$$

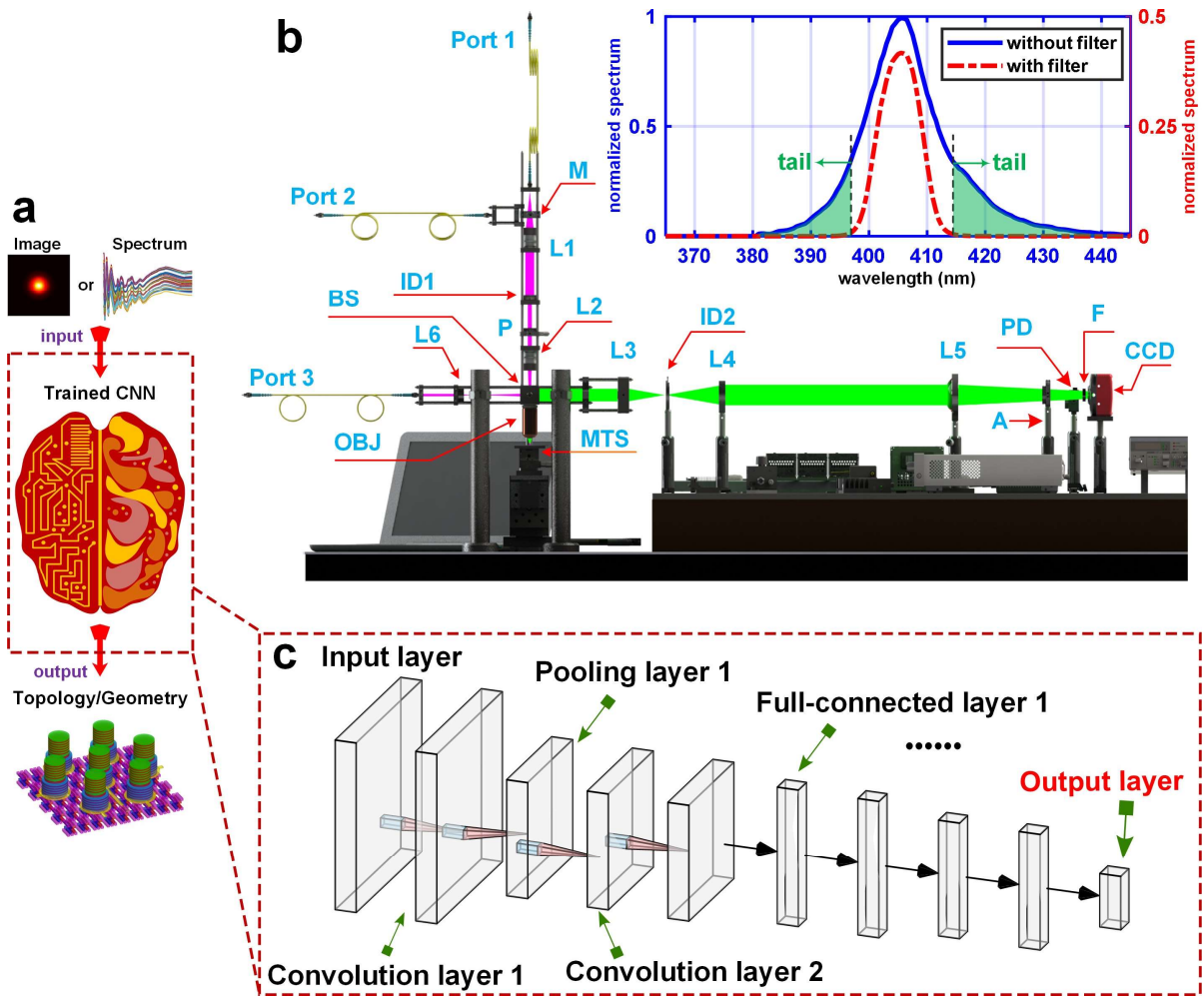

Figure 1: Schematic of the proposed optical deep learning nano-profilometry. (a) Simplified flowchart. (b) Schematic of the in-house BFM. (c) Architecture of the CNN model. The inset at the top right corner of (b) presents the spectrum of the 405-nm LED source with and without the bandpass filter. M, mirror; L, lens; P, polarizer; A, analyzer; PD, photodetector; OBJ, objective; ID, iris diaphragm; MTS, motorized translation stage; F, filter; CCD, charge-coupled device. Ports 1 and 2 connect the 405-nm LED and the 532-nm laser sources, respectively. Port 3 is used to capture the spectrum of the source. The retractable rotating mirror M is utilized to guarantee only one source is selected in the measurement. Only the 405-nm LED source is used here. The filter in the imaging space significantly filters the tails of the LED spectrum.
3

Here $\hat{y}_i^{(n)}$ is the *i*-th reconstructed dimension of the *n*-th data sample and $y_i^{(n)}$ is the corresponding actual dimension. *N* denotes the size of the training data set whereas *M* is the number of geometrical dimensions to be extracted. We introduce $\omega_{yi}$ here as the weight term associated with the *i*-th dimension. We call this cost function the weighted mean squared error (WMSE). The weight term $\omega_{yi}$ offers an extra degree of freedom in controlling the mapping dynamics of the network and allows us to treat each geometrical dimension with its own criticality. Depending on the measurement system used, the input to the network can be a 1-D, 2-D, or 3-D tensor, while the output is a vector whose elements correspond to the dimensions to be estimated. For training, the network parameters are randomly initialized following the method proposed in [43]. We then repeatedly sample a batch (batch size = 1024) of training data and perform a gradient based optimization on Θ. We use the Adam optimizer [44] with $\beta_1 = 0.9$, $\beta_2 = 0.999$, and the learning rate initially set to $1 \times 10^{-3}$ and gradually decaying to $1 \times 10^{-6}$. The CNN model and the training are implemented with the Keras library in Python. The computing platform consists of an Intel Core i7-7700K CPU and a Nvidia GTX-1080 GPU. We have also made our implementation and pre-trained models open-source on our GitHub project page [45].

## 2.3 Measurement systems

We used two types of optical systems, i.e., BFM and ellipsometry, to validate the proposed framework in this paper. The first instrument is an in-house epi-illumination microscope equipped with a fiber-coupled 405-nm light emitting diode (LED) source (M405FP1, Thorlabs Inc.) and a CCD camera (C4742-80-12AG, Hamamatsu Inc.). The magnification of the BFM system is 106.7×. An insertable bandpass filter (FB405-10, Thorlabs Inc.) centered at 405 nm and with a 10-nm full width at half maximum (FWHM) is used in front of the CCD to narrow down the spectrum for accurate model mapping. A Zeiss plan-apochromat 20×/0.8 numerical aperture (NA) objective lens is used together with the 5.33×4-f lens system (L4 and L5). Fourier optics is adequately accurate for the generation of training sets for CNN. The detailed configuration of the BFM is illustrated in Fig. 1b. The second system is an ellipsometer (ME-L Mueller matrix ellipsometer, Wuhan Eoptics Technology Co., Wuhan, China) that is conventionally utilized to analyze the optical constants of thin films within the wavelength range of 193–1690 nm. The instrument can also be operated in incident angle-resolved and azimuthal angle-resolved modes. In the Results and Discussion section, we will discuss in detail the operation modes we used in the ellipsometry examples.

## 2.4 Sample preparation and measurement schemes

The samples under investigation include the silicon fins on a NIST RM 8820 artifact [32], a nanowire array fabricated by nanoimprinting, and a DRAM transistor consisting of three-layer nanostructures [$Si_3N_4$, $SiO_2$, and (100)-orientation Si, from top to bottom] with several features that are smaller than 10 nm. Figure 2d shows the SEM image of the investigated pattern (pattern G) on the NIST artifact. The geometrical dimensions of the silicon fins and the material constitutions can be found in [32]. The nanoimprinted wires have a similar shape as that of the NIST artifact, but with a transverse dimension that is twice as small and a much taller height. See Fig. 3b. The DRAM transistor was manufactured by a standard 45-nm node process. Because of the difference in etching anisotropy between $Si_3N_4$ and Si, the top and bottom trapezoids have different sidewall angles [see Fig. 4b obtained by a TEM (TE20, TEM.FEI Co.)]. This asymmetry needs to be taken into consideration in the simulation when generating training data for the CNN.

## 3 Results and Discussion

### 3.1 BFM-based nanometrology

The first optical modality we used with the proposed framework is BFM. The measurand under consideration is a NIST RM 8820 artifact, with an array of short lines characterized by three geometrical dimensions, namely the bottom width (BCD), the height (HGT), and sidewall angle (SWA) (see Fig. 2d for a cross-sectional view). For data generation, we take 2,400 random samples from $BCD \sim U(1000, 1100)$ nm, $HGT \sim U(80, 100)$ nm, and $SWA \sim U(82, 90)°$, where $U(a, b)$ refers to the uniform probability distribution on the interval $[a, b]$. We divide this into groups of 2000, 200, and 200 samples for the training set, validation set, and test set, respectively. We use Fourier optics-based simulation at $\lambda = 405$ nm to obtain the microscopy image of the samples. Note that the spatial resolution of the simulation must match that of the BFM measurement system, which in our case is 101×101. We take the average of all the horizontal slices of the experimental image to alleviate perturbations due to measurement noise, dust on the sample, and imperfect sources, and use the average slice as the input to the CNN.

The CNN is trained for 3000 epochs in around 3 minutes; see Fig. S1a in Supplement 1 for the training curve. After training, we first test its performance on the simulation data. Figures 2a–c show the results on the test set, where the *x*- and *y*-axes are the nominal and reconstructed dimensions, respectively. The blue dots correspond to the 200 samples in the test set. The red lines serve as a guide to the eye for perfect reconstruction. We also inspect the results quantitatively with the average bias $B$ and standard deviation $\sigma$, defined as the mean and standard deviation of $|D_{real} - D_{recon}|$, where $D_{real}$ and $D_{recon}$ are the actual and reconstructed dimensions, respectively. The average bias and standard deviation are at sub-nanometer and sub-degree scales, indicating the good reconstruction accuracy of the proposed method; see the values in the insets of Figs. 2a–c. Note that the longitudinal dimensions (*HGT* and *SWA*) cannot be directly determined from a single-shot optical image in conventional BFM. However, we are able to recover these dimensions here because the intensity distribution of the BFM image is a nonlinear function of all the geometrical dimensions of the nanostructures. In particular, the BFM image contains the diffraction patterns from light scattering at the edges of the nanostructures and our CNN method can identify these patterns in the images with sufficient training. We also want to point out that once trained, the parameter extraction (inference) step is a single feed-forward process in the neural network computation, which is very fast. It takes less than 10 milliseconds for the 200 samples in the test set.

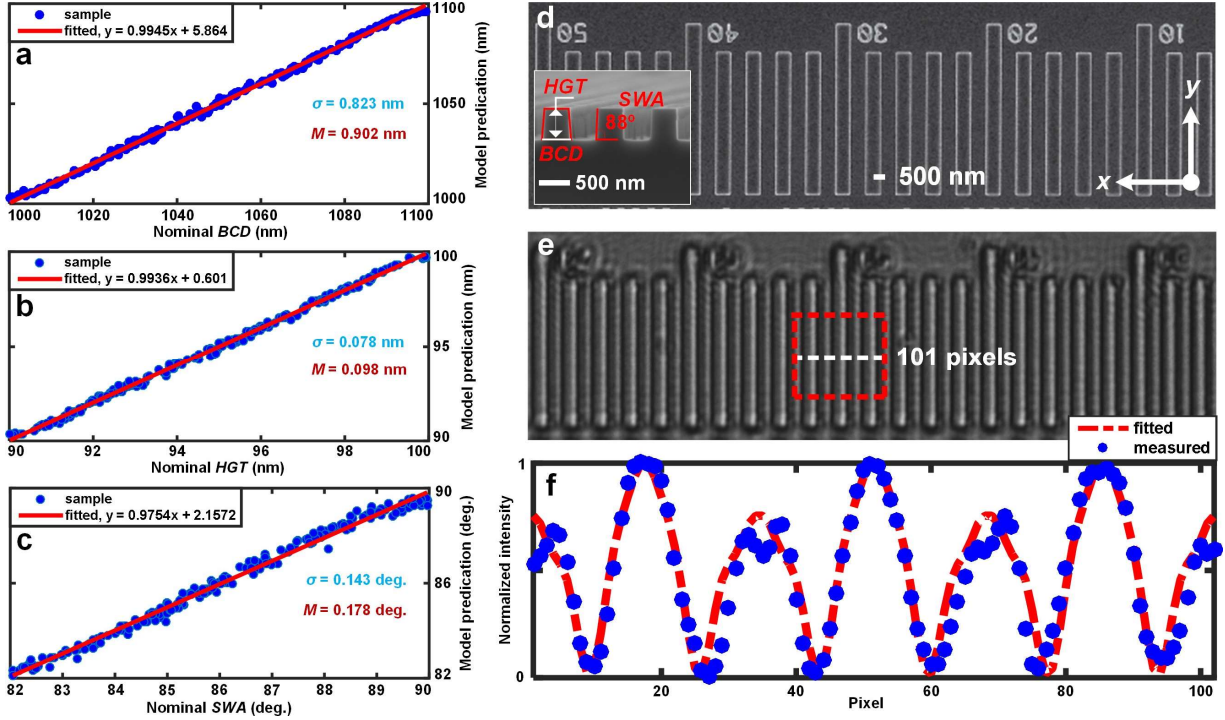

Figure 2: Reconstruction for the NIST RM 8820 pattern. (a–c) The extracted geometrical dimensions for *BCD*, *HGT*, and *SWA* for 200 testing samples (simulation). The average bias ($M$) and the standard derivation of the bias ($\sigma$) are marked in each plot. (d), (e) SEM and BFM top-view images of the investigated area. The SEM images are adapted from [32]. (f) Cross-sectional intensity curve across the three device lines within the red cropped window of (e) for the measured image and the image calculated from the CNN output. The measured curve in (f) averages all of the 101-pixel wide horizontal slices, such as the white dotted line in (e), for different y-positions. The inset in (d) is the SEM side-view image of a similar sample showing the three dimensions: *BCD*, *HGT*, and *SWA*.

We can now evaluate the performance of our CNN model for retrieving dimensions from experimental BFM images. The measured BFM image contains three periods of lines, which are chosen in such a way that the sample stage induced tilting is minimal, as determined by the regularized pseudo-phase imaging method [46]. For the actual measurand and the measured BFM image, the CNN-reconstruction has an output of $BCD = 1.00$ μm, $HGT = 91.43$ nm and $SWA = 90°$ (see one of the measured images in Fig. 2e as an example). Compared with the NIST [32] measured dimensions of 1.00 μm, 97.3 nm and 88° (the *SWA* was measured from a similar silicon nanowire but with a larger height-width-ratio [32]), our model has a bias of 5.9 nm and 2.0° for *HGT* and *SWA*, respectively, and a perfect reconstruction for *BCD*. The likely sources for the mismatch are systematic BFM measurement errors, e.g., our inability to place the sample at the focal plane with an accuracy better than 50 nm, and random BFM measurement noise. Using the reconstructed dimensions, we simulate the expected image compare it with the BFM image of the actual measurand (horizontal slice only) in Fig. 2f.

## 3.2 Ellipsometry-based nanometrology

The second modality we worked with is ellipsometry. Ellipsometry measures the polarization-sensitive scattering spectrum at either multi-wavelengths or multi-angles (incident or azimuthal angle), which is completely different from BFM in terms of the measurands and operation modes. We use an ellipsometer operating in the specular reflection mode (see Fig. 3a). The incident angle $\theta$ and the azimuthal angle $\varphi$ can be tuned continuously via a rotation stage and a single-axis rotation arm (not shown). The source can be either a laser or a broadband source, depending on the detector in use. We first operate the ellipsometer in angle-resolved mode: we scan both $\theta$ and $\varphi$ and measure the scattering spectrum versus wavelength for each angle combination. Here, we select five incident angles (in the range 45° – 65°; 5° increment) and nineteen azimuthal angles (in the range 0° – 90°; 5° increment) to obtain the 5×19×15 spectra, where the last dimension of size 15 comes from the 15 elements of the normalized Mueller matrix; see the visualization in Fig. 3c. Data is simultaneously collected at ten wavelengths from 350 nm to 800 nm with a 50-nm increment. The 4-D data at each wavelength is used to train an individual CNN. Because the datum at different wavelengths are measured independently by different pixels in the spectrometer of the ellipsometer, the measured data points in the spectrum are uncorrelated. Thus, we can average the output dimensions of the CNN models at various wavelengths to improve the overall accuracy. The measurand under investigation is a silicon nanowire, characterized by three geometrical dimensions, *TCD*, *HGT* and *SWA* (see Fig. 3b). We sample from *TCD*~$U$(300, 400) nm, *HGT*~$U$(430, 530) nm, and *SWA*~$U$(82, 90)°, to get 3200 samples and use the RCWA algorithm to obtain the corresponding scattering spectra. The data are divided into 2500 for training, 500 for validation, and 200 for test. For our ten CNN models, each model is trained for 2500 epochs in around 4 minutes; see Fig. S1b in Supplement 1 for the training curve of a typical model.

The results on the simulation data are presented in Fig. 3d, where we have three sets of box plots showing the relative percentage errors for the reconstructed *TCD*, *HGT*, and *SWA* on the test set for the ten models trained at different wavelengths. We are able to achieve less than 0.3% errors with each individual model. Figure 3e shows the experimental results. We measure the nanowire at the same ten wavelengths and feed the scattering data to the corresponding CNNs. The dimensions measured by SEM (dotted lines) are also shown within each sub-figure. The dimensions reconstructed by CNN at each single wavelength are all very close to the SEM measured values of *TCD* = 350 nm, *HGT* = 472 nm, and *SWA* = 88°, while the average reconstruction over the ten models is *TCD* = 350.4 nm, *HGT* = 471.6 nm, and *SWA* = 87.4°, corresponding to an accuracy of $\Delta_{TCD}$ = 0.4 nm (0.11%), $\Delta_{HtT}$ = 0.4 nm (0.08%), and $\Delta_{SWA}$ = 0.6° (0.68%). Here, we should mention that during the measurement, there are uncertainties caused by the positioning errors of the mechanical components (such as the RS and the single-axis rotation arm) and the non-zero spectral bandwidth of the scattering signal captured by each pixel in the spectrometer (the detector), which could degrade the reconstruction accuracy of the CNN. We expect the reconstruction accuracy of the proposed method can be further improved by reducing measurement errors, which can be achieved by using opto-electronic components with better performance and better system calibration using automatic platforms.

To further validate the generality of the proposed method, we fix the incident and azimuthal angles at 65° and 0°, respectively, while only capturing the wavelength-resolved scattering spectrum. The source has an operating wavelength in the range of 200-800 nm with a 10-nm increment, and we use the resulting 61× 15 spectra as the input to the CNN. We investigate a DRAM transistor, whose geometrical profile consists of three parameters $D_1$, $H_1$, $H_2$, $H_3$, $SW\ A_1$, and $SW\ A_2$; see the schematic and the TEM measured cross-section in Figs. 4a and 4b, respectively. We sample from $D_1$~$U$(50, 100) nm, $H_1$~$U$(110, 160) nm, $H_2$~$U$(3, 23) nm, $H_3$~$U$(110, 160) nm, $SWA_1$~$U$(82, 90)°, and $SWA_2\ U$ (82, 90)°, to get 3,200 samples and use the RCWA algorithm to obtain the scattering spectra. Among all the data, 2500 are used for training, 500 are used for model selection, and 200 for test.

The CNN is trained for 2,500 epochs in around 4 minutes; see Fig. S1c in Supplement 1 for the training curve. Figures 4c–e show the performance of the CNN on the test set of the simulation data for $D_1$, $H_2$, and $SWA_1$, which are the most critical dimensions governing the performance of the DRAM transistor. We quantify the accuracy of our predictions by calculating the intervals containing 68% and 95% of the reconstructed dimensions from their nominal values, indicated by the orange and black lines in these plots. Similar to Figs. 2a–c, the horizontal and vertical axes correspond to the nominal and reconstructed dimensions, respectively, whereas each dot represents a data point in the test set and the red lines serve as a guide to the eye for perfect reconstruction. We can see the high reconstruction accuracy of the CNN model.

We can further improve the reconstruction accuracy for an individual dimension by adjusting its associated weight in the WMSE loss function. Here we demonstrate this process on the thickness of the $SiO_2$ layer denoted as $H_2$. There are two major reasons for this adjustment. First, without any special treatment, the prediction accuracy is worst for $H_2$ among all six dimensions. This is because the scattering signature is very insensitive to $H_2$. Second, $H_2$ is the thickness of the central layer and governs the leakage current of the DRAM unit and thus has a significant effect on the device performance. Therefore, accurately determining its value is critical. The optimal value of $\omega_{H2}$ is determined by performing hyper-parameter tuning on the validation set. The improved accuracy can be observed by comparing Figs. 4d and 4g, where the relative error reduces from 4.38% to 1.66%. Note that to achieve this improvement, the reconstruction accuracy for other dimensions dropped slightly, as shown in Figs. 4f and 4h. Thus, there is a trade-off to make.

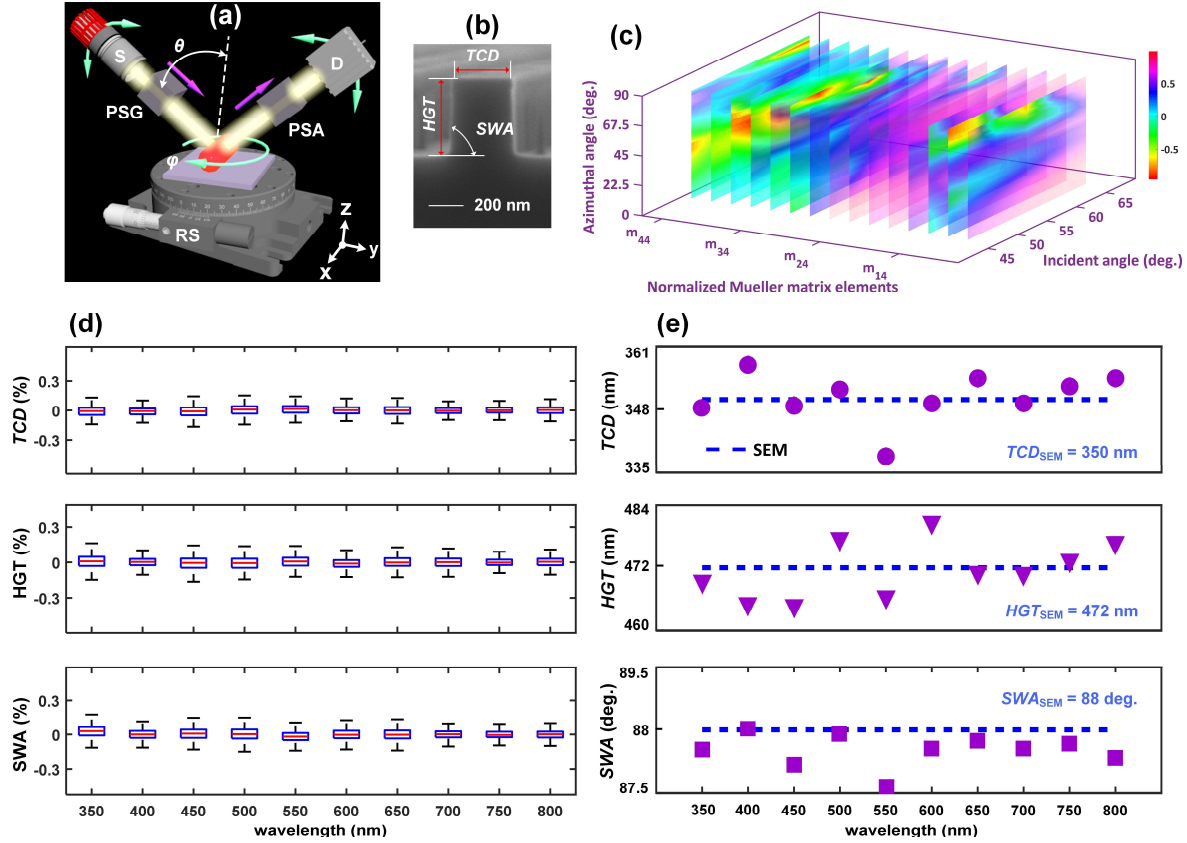

Figure 3: Dimension reconstruction for a nanowire array with CNN-based ellipsometry operating in the angle-resolved mode. (a) The simplified schematic of the ellipsometer operating in the specular reflection mode. (b) The SEM cross-section image of the nanowire. (c) The 4D plot of a representative 5×19×15 spectrum. (d) Box plots of the relative percentage error on the test set of the simulation data. (e) The reconstructed dimensions of *TCD*, *HGT*, and *SWA* of the silicon nanowire. S, source; D, detector; RS, rotation stage; PSG, polarization state generator; PSA, polarization state analyzer. Because the SEM measurement does not depend on wavelength, we use the dashed line in (e) to indicate that they are obtained from a local position of the nanowire.

We next consider the reconstruction with measurement data. We select four dies on a 12-inch wafer to take the die-to-die variations into account. Each die is measured thirty times and we use the averaged spectrum to mitigate random perturbations in the measurement. As can be seen in Figs. 5b–g, we get an improved reconstruction for $H_2$ and $H_3$ when $\omega_{H2}$ is larger, whereas the accuracy for the other dimensions slightly degrades. Nonetheless, all the reconstructed dimensions are adequately accurate for good predictions of DRAM performance. The difference in the reconstructed values obtained for different dies is partly due to the inevitable fabrication errors (e.g., patterning uniformity, line-edge roughness, and line-width roughness) and measurement errors (e.g., random measurement noise as well as positioning error when translating the wafer horizontally). We expect that the reconstruction accuracy through the proposed method can be improved by further reducing the measurement errors.

## 4  Conclusion

In this paper, we introduced a CNN-assisted nano-metrology framework for the reconstruction of deep-subwavelength geometrical profiles of nanostructures. The proposed framework works with diffraction-limited optical modalities such as the bright-field microscopy and the optical ellipsometry, and is non-contact, non-destructive and fast in run-time. We demonstrated the effectiveness and generality of the proposed method by the successful reconstruction of profiles (including sub-10-nm dimensions) of various nanostructures, which are widely seen in fields like semiconductor industry, integrated photonics, metamaterials, one-dimensional photonic crystals, biosensors, and neuromorphic chips, at a sub-nanometer scale accuracy using single-shot microscopy images or scattering spectra. Moreover, we showed that the reconstruction accuracy with respect to individual dimensions is adjustable, enabling us to investigate the most critical dimensions that govern the performance of nanoscale devices. The proposed nanometrology framework is built upon the power of neural networks in representing the highly nonlinear mapping between the scattering information and the geometrical dimensions of the measurands and is expected to work with other diffraction-limited optical modalities.

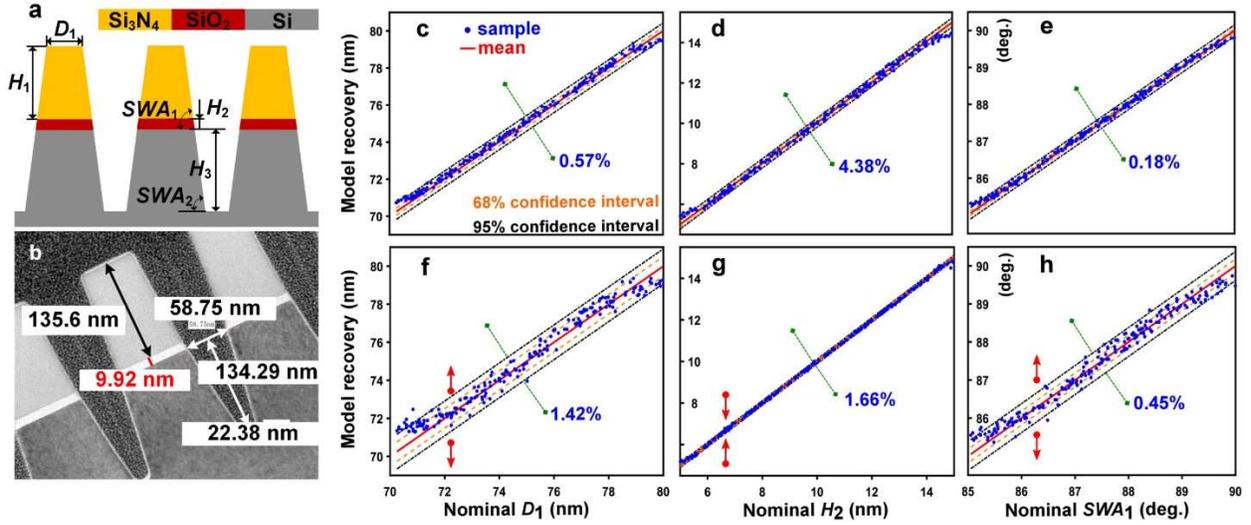

Figure 4: Dimension reconstruction of a DRAM array with CNN-based ellipsometry operating in wavelength-resolved mode. (a),(b) Geometrical model and TEM image of the DRAM array, showing that the critical dimension (marked in red) is smaller than 10 nm. (c–e) The reconstructed dimensions of $D_1$, $H_2$, and $SWA_1$ on the test data when all $\omega_i = 1$. The orange and black lines denote the 68% and 95% confidence intervals, respectively. (f–h) The reconstruction results on the test data when $\omega_{H2} = 10$. The thickness of the 95% confidence interval bands (normalized by the nominal dimensions) are reported in each sub-figure. The red arrows in (f–h) denote the change in the confidence interval bands with respect to those from (c–e) due to the increase in $\omega_{H2}$.

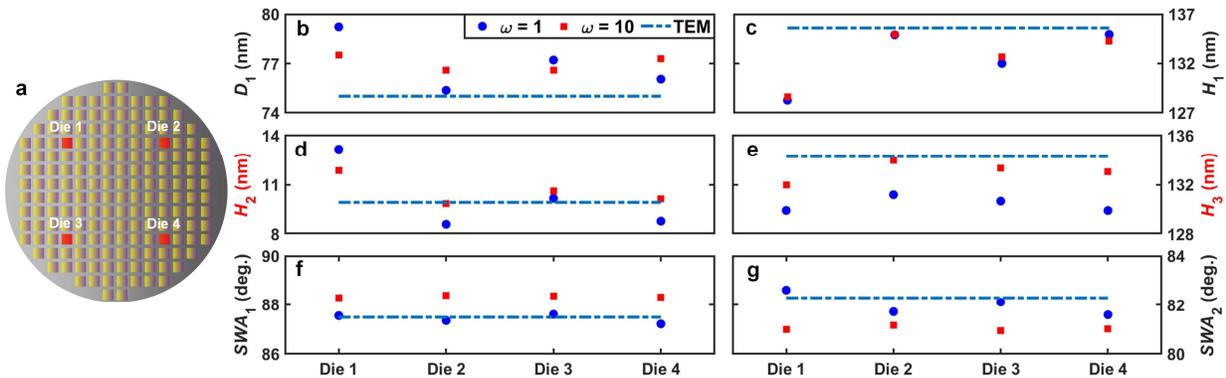

Figure 5: Experimental reconstruction of four DRAM transistors on a 12" wafer. (a) Schematic of the 12" wafer and the investigated four dies. (b–g) The averaged reconstruction dimensions of $D_1$, $H_1$, $H_2$, $H_3$, $SWA_1$, and $SWA_2$ for 30 repeated measurements within each die. The TEM measured values are represented by the dark cyan dash-dotted line. Because the TEM measurement is conducted only on a single DRAM transistor, we use the dash-dotted line to connect the four values in (b–g) to indicate that they are obtained from the same TEM measurement.

## Acknowledgments

We acknowledge Cisco for access to its Arcetri cluster. J. Zhu, S. Purandare, and L. L. Goddard acknowledge funding from Cisco, UIUC, and ZJU-UIUC; Y. Liu and J.-M. Jin acknowledge funding from Cisco; S. Liu acknowledges funding from NSFC.

# Supplementary Material

**CNN Model Architecture**

Our CNN models are adapted from the VGGNet architecture and are modified for critical dimension extraction with the weighted mean-square-error (WMSE) regression loss. We use the same basic model but with different parameters for our three examples, due to the difference in input size. For the NIST 8820 artifact example, we use bright-field microscopy (BFM) to get a 2-D image of the measurand. We crop the image to a size of 101×101. We then take the average of the 101 horizontal slices of the image, resulting in an input of size 101×1. For the nanowire example, we use angle-resolved ellipsometry to get the scattering spectra of the measurand, resulting in an input size of 5×19×15, for the five incident angles, nineteen azimuthal angles, and fifteen elements of the normalized Muller matrix. In the DRAM case, we use wavelength-resolved ellipsometry with sixty-one wavelengths, resulting in an input of size 61×15 for the sixty-one wavelengths and fifteen elmenets of the normalized Muller matrix. We use convolution of kernel size 2 and max pooling of size 2 along each input dimension. We normalize the input and output to be within range [0, 1], and use the rectified linear unit (ReLU) activation function defined as $y(x) = \max(0, x)$ with convolution layers and the first fully-connected layer, and the sigmoid activation with the last fully-connected layer. Table S1 lists the details of the CNN models.

Table S1: Architectures of the CNN models used in this paper.

| NIST 8820 | Nanowire | DRAM |
|---|---|---|
| Input: 101 × 1 | Input: 5 × 19 × 15 | Input: 61 × 15 |
| Conv: (100, 2, ReLU) | Conv: (100, 2 × 2 × 1, ReLU) | Conv: (100, 2 × 2, ReLU) |
| Batch Norm | Batch Norm | Batch Norm |
| Conv: (50, 2, ReLU) | Conv: (50, 2 × 2 × 1, ReLU) | Conv: (50, 2 × 2, ReLU) |
| Batch Norm | Batch Norm | Batch Norm |
| Max Pooling (2) | Max Pooling (2 × 2 × 2) | Max Pooling (2 × 2) |
| Drouput (0.9) | Drouput (0.9) | Drouput (0.9) |
| FC (50, ReLU) | FC (50, ReLU) | FC (50, ReLU) |
| Batch Norm | Batch Norm | Batch Norm |
| Drouput (0.9) | Drouput (0.9) | Drouput (0.9) |
| FC (3, sigmoid) | FC (3, sigmoid) | FC (6, sigmoid) |

**CNN Training**

For training, we use the Xavier initialization and the Adam optimizer, with the batch size chosen to be 1024, and $\beta_1 = 0.9$, $\beta_2 = 0.999$. The learning rate is set initially to $1\times10^{-3}$, and decays by a factor of 5 whenever the validation loss hits a plateau, until it reaches $1\times10^{-6}$. We also checkpoint the network parameters every 10 epochs and use early stopping to terminate the training once the validation loss is not decreasing for over 500 epochs. The training curves for the three examples can be found in Fig. S1. The training is done on a single desktop computer with an Intel Core i7-7700K CPU and NVIDIA GTX-1080 GPU.

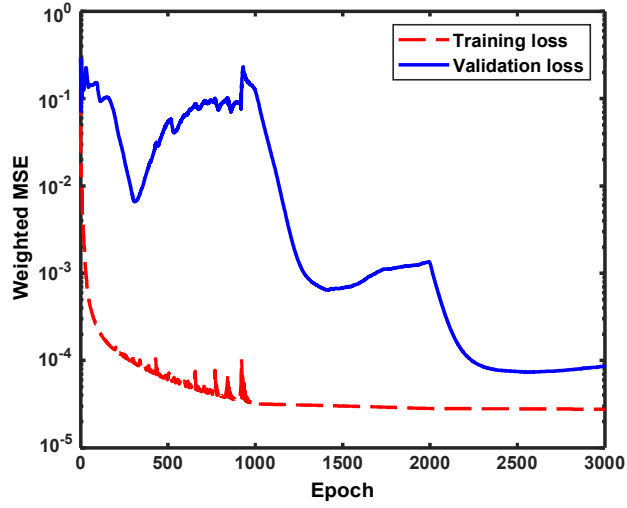

(a) The NIST 8820 artifact

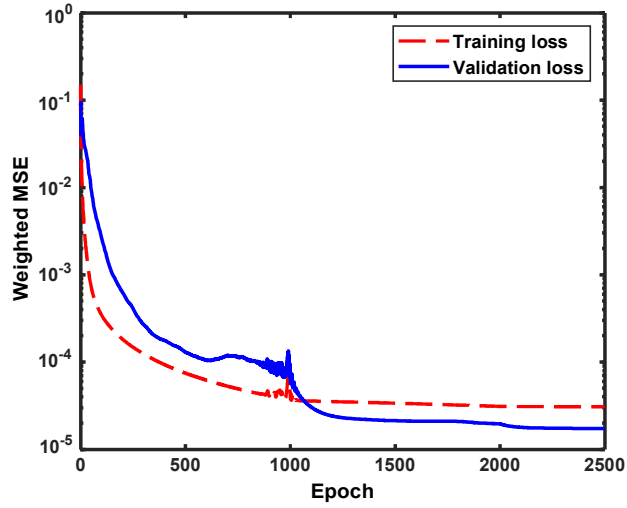

(b) The nanowire array.

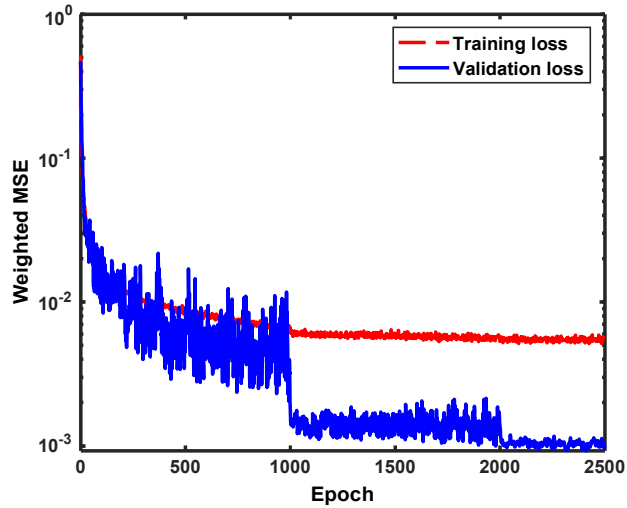

(c) The DRAM transistor.

Figure S1: Training curves for the CNNs in three examples of this paper.